# Study of Iterative Detection and Decoding with Log-Likelihood Ratio Based Access Point Selection for Cell-Free Networks

Roberto B. Di Renna, *Member, IEEE,* and Rodrigo C. de Lamare, *Senior Member, IEEE*

*Abstract*—This paper proposes an iterative detection and decoding (IDD) scheme and an approach to improve the selection of access points (APs) in uplink cell-free massive multiple-antenna systems. A cost-effective scheme for selection of APs based on local log-likelihood ratios (LLRs) is developed that provides sufficient statistics to the central processing unit and selects which APs should be considered for each user. Numerical results show that the proposed IDD scheme works very well and the proposed LLRs-based approach to select APs outperforms the existing techniques in terms of bit error rate and spectral efficiency while requiring a comparable fronthaul load.

## I. INTRODUCTION

The high demand for data traffic and the need to serve a massive number of users in the network present significant challenges for the next generation of wireless communication systems [1], [2], [3]. In order to address these problems, maximize the spectral efficiency (SE) of each user and provide more uniform performance across users, cell-free massive multiple-input multiple-output (MIMO) systems has been advocated in recent years. The main idea is to employ a large number of distributed access points (APs) to support a massive number of users in the same time-frequency resource block [4]. Each access point (AP) is connected to a central processing unit (CPU) which is responsible for coordinating and processing the signals of the users. Since all APs are connected to a CPU via fronthaul links, there are actually no cell boundaries [5]. Furthermore, cell-free MIMO systems present many technical challenges and limitations that include scalability, power consumption, deployment of APs, fronthaul signaling, extra spectrum if a wireless fronthaul is used and resource allocation, which call for innovative technical solutions. In particular, to achieve the expected theoretical gains under an ideal situation, each UE should be served by all APs in the network [6], [7], [8]. However, this is not scalable with increased network size due to huge fronthaul signaling and computational requirements. Moreover, the distributed location and the delay spread of the APs, and the associated signal propagation latency will limit the APs involved in cell-free MIMO systems. In this context, AP selection techniques, which correspond to selecting a group of APs according to a desired criterion, are key to reducing fronthaul signaling, computational costs and latency [9].

Previous works on AP and relay selection [10] are based on channel statistics. Considering the geographic distribution of the users, the work in [11] proposes an AP selection scheme based on the largest large-scale fading (LLSF) and the minimum geographic distance between APs and users. In order to improve the system performance of [11], the work in [12] adopted the signal-to-interference-plus-noise ratio (SINR) to compute the largest effective channel gain (LECG) as a criterion to select the APs of user $k$. In addition, the work of Vu *et al.* in [13] introduced an AP selection scheme based on sum-rate optimization. The work in [14] devised an AP selection scheme based on a downlink power minimization problem. The study in [15] sends local LLRs on the fronthaul link instead of the soft-detected symbols, which considerably increases the computational cost and uses the LLSF AP selection scheme. Related works studied energy efficiency optimization and power allocation problems to increase the spectral efficiency [16], [17], [18], but they do not use the already available local soft detected symbols to improve the AP selection and the bit error rate (BER).

In this work, we propose an iterative detection and decoding (IDD) scheme that operates in two stages in the uplink of cell-free massive MIMO systems and exploits the cooperation between the APs and the CPU. Based on the large-scale fading parameters, in the first stage each AP locally estimates the channels and applies a linear receive filter to obtain local estimates of the user symbols. These estimates are then organized at the CPU where they are linearly processed prior to detection. In the second stage at the CPU, only the channel statistics are used as we assume for simplicity that the pilot signals are not shared over the fronthaul links. In the proposed IDD scheme, we develop an approach that uses the already available local soft estimates of the symbols to compute bit log-likelihood ratios (LLRs) before sharing the local estimates (channel and data) on the fronthaul links. The iterative processing with the exchange of LLRs between the detectors and the decoders is performed at the CPU. Unlike previous works [5], [19], [17], [12], [20] that consider that each user is supported by a subset of APs based on LLSF or channel gain metrics, we devise an approach that uses the LLRs to select users served by each AP, without increasing the fronthaul signaling load required by recent techniques. Simulations show that including the local LLR computation on the decision procedure of the subset of APs benefits the system performance in BER and cumulative distribution function. A study of the fronthaul signaling and the computational complexity is also carried out.

This paper is organized in the following sections: Section II details the system model and the channel estimation. The proposed decentralized detection and IDD scheme are presented in Section III, where the LLRs computation and the AP selection scheme are detailed. Section IV analyzes the fronthaul signaling load and the computational cost required for the proposed techniques as compared with existing approaches. Simulations are provided in Section V, whereas the conclusions are drawn in Section VI.

## II. SYSTEM MODEL AND CHANNEL ESTIMATION

We consider the uplink of a cell-free massive MIMO system

The authors are with the, Pontifical Catholic University of Rio de Janeiro, Rio de Janeiro 22453-900, Brazil (e-mail: delamare@puc-rio.br). This work was supported by CNPq, FAPERJ and FAPESP.

that consists of $L$ APs, each equipped with $N$ antennas, that cover an area with $K$ single-antenna users. The APs are connected to the nearest CPU via fronthaul links, where the symbol decoding is performed. This architecture permits coherent transmission and reception to the users in the entire coverage area and fits in the massive MIMO category under the assumptions that $L$ and $K$ are large, and $L \gg K$ [4].

The channel follows the block fading model, is constant over a transmission frame of duration $\tau_c$ and changes independently at each coherence block. The channel fading parameters between the $k$th user and the $l$th AP are obtained from a circularly symmetric complex Gaussian distribution with zero mean and spatial correlation matrix $\mathbf{R}_{kl} \in \mathbb{C}^{N \times N}$, that is, $\mathbf{h}_{kl} \sim \mathcal{N}_\mathbb{C}(\mathbf{0}, \mathbf{R}_{kl})$. The Gaussian distribution models the small-scale fading whereas large-scale fading, including geometric pathloss, shadowing, antenna gains and spatial channel correlation [5]. In this study, this parameter is described as $\beta_{kl} = \operatorname{tr}(\mathbf{R}_{kl})/N$. Since this scenario implies that the APs are geographically distributed, we can assume that the channel vectors of each AP are independent and identically distributed (i.i.d.), therefore $\mathbb{E}\left\{\mathbf{h}_{kn}(\mathbf{h}_{kl})^\mathsf{H}\right\} = \mathbf{0}$. Thus, we consider that the channels of each user are also i.i.d. and the spatial correlation matrices are available in the entire network [6].

In the uplink scenario under study, each coherence block is divided into $\tau_p$ channel uses for pilots and $\tau_u$ for data such that $\tau_c = \tau_p + \tau_u$. For channel estimation, we use $\tau_p$ mutually orthogonal $\tau_p$-length pilot signals, where $\|\phi_t\|^2 = \tau_p$. Even though we consider a large number of APs for each user, we are interested in the case of $K > \tau_p$, which results in pilot sequence sharing among users. In order to obtain the channel estimates, we employ the minimum mean square error (MMSE) estimator. Thus, each AP correlates the received signal with the associated pilot sequence, as given by

$$\dot{\mathbf{y}}_{tl} = \sum_{i \in \mathcal{P}_t} \sqrt{p_i \tau_p} \mathbf{h}_{il} + \mathbf{v}_{tl}, \ \in \mathbb{C}^{N \times 1} \tag{1}$$

where $p_i$ is the transmit power for each sample of the pilot sequence of the $i$th user, $\mathcal{P}_t$ is the subset of users assigned to pilot $t$ and the resulting noise $\mathbf{v}_{tl}$ can be approximated as $\mathbf{v}_{tl} \sim \mathcal{N}_c(0, \sigma_v^2 \mathbf{I}_N)$ where $\sigma_v^2$ is the noise power. The MMSE estimate of $\mathbf{h}_{kl}$ for $k \in \mathcal{P}_t$ is given by

$$\hat{\mathbf{h}}_{kl} = \sqrt{p_k \tau_p} \mathbf{R}_{kl} \boldsymbol{\Phi}_{tl}^{-1} \dot{\mathbf{y}}_{tl}, \ \in \mathbb{C}^{N \times 1} \tag{2}$$

where $\boldsymbol{\Phi}_{tl} = \mathbb{E}\left\{\dot{\mathbf{y}}_{tl} \dot{\mathbf{y}}_{tl}^\mathsf{H}\right\} = \sum_{i \in \mathcal{P}_t} \tau_p p_i \mathbf{R}_{il} + \sigma_v^2 \mathbf{I}_N$ is the correlation matrix of the received signal in (1). The channel state information (CSI) $\hat{\mathbf{h}}_{kl}$ and the estimation error $\tilde{\mathbf{h}}_{kl} = \mathbf{h}_{kl} - \hat{\mathbf{h}}_{kl}$ are statistically independent vectors distributed as $\hat{\mathbf{h}}_{kl} \sim \mathcal{N}_c\left(\mathbf{0}_N, p_k \tau_p \mathbf{R}_{kl} \boldsymbol{\Phi}_{tl}^{-1} \mathbf{R}_{kl}\right)$ and $\tilde{\mathbf{h}}_{kl} \sim \mathcal{N}_c\left(\mathbf{0}_N, \mathbf{C}_{kl}\right)$ with $\mathbf{C}_{kl} = \mathbb{E}\left\{\tilde{\mathbf{h}}_{kl} \tilde{\mathbf{h}}_{kl}^\mathsf{H}\right\} = \mathbf{R}_{kl} - p_k \tau_p \mathbf{R}_{kl} \boldsymbol{\Phi}_{tl}^{-1} \mathbf{R}_{kl}$.

## III. Proposed IDD Scheme and Decentralized Detection

In this section, we describe the symbol transmission phase, the detection steps, the LLR computation and the AP selection of the proposed IDD scheme. Each AP collects the received signals, locally estimates the channels, performs soft symbol estimates and then sends the local soft symbol estimates of the signals to the CPU for detection and decoding. In our IDD scheme that is a distributed adaptation of the soft interference cancellation (IC) scheme of [21], we compute local bit LLRs to improve AP selection and transmit only the list of supported users of each AP and the channel statistics over the fronthaul links. After channel estimation, the signal received at the $l$th AP in the data transmission phase is described by

$$\mathbf{y}_l = \sum_{i=1}^K \sqrt{\eta_i} \mathbf{h}_{il} x_i + \mathbf{v}_l, \ \in \mathbb{C}^{N \times 1} \tag{3}$$

where $\eta_i$ and $x_i$ denote the uplink transmit power and the bits of the $k$th user mapped into a modulation alphabet, such as quadrature phase shift keying (QPSK), in the symbol interval. At this cooperation level, each AP produces local estimates of the symbols that are then transmitted to the CPU for detection and decoding. In order to obtain the local soft symbol estimate $\hat{x}_{kl}$, let $\mathbf{w}_{kl}$ be the local receive filter of the $l$th AP that is selected for user $k$, which yields

$$\hat{x}_{kl} \triangleq d_{kl} \mathbf{w}_{kl}^\mathsf{H} \mathbf{y}_l \tag{4}$$

$$\triangleq d_{kl} \mathbf{w}_{kl}^\mathsf{H} \sqrt{\eta_k} \mathbf{h}_{kl} x_k + d_{kl} \mathbf{w}_{kl}^\mathsf{H} \sum_{i=1, i \neq k}^K \sqrt{\eta_i} \mathbf{h}_{il} x_i + d_{kl} \mathbf{w}_{kl}^\mathsf{H} \mathbf{v}_l,$$

where $d_{kl}$ is a binary scalar that indicates if all antennas of the $l$th AP supports the $k$th user.

Using the CSI obtained in (2), the linear receive filter that minimizes the MSE is described by

$$\mathbf{w}_{kl} \triangleq \eta_k \left( \sum_{i \in \mathcal{D}_l} \eta_i \left( \hat{\mathbf{h}}_{il} \hat{\mathbf{h}}_{il}^\mathsf{H} + \mathbf{C}_{kl} \right) + \sigma_v^2 \mathbf{I}_N \right)^\dagger \hat{\mathbf{h}}_{kl}, \tag{5}$$

where $(\cdot)^\dagger$ refers to the pseudoinverse, $\mathcal{D}_l$ is the set of users served by AP $l$ which provides a more scalable solution than the standard form of cell-free MIMO systems, where an $LN \times LN$ matrix must be inverted [5], [6].

The CPU receives the locally obtained soft estimates of the symbols and performs detection and decoding by using the large-scale fading decoding (LSFD) weighted signal approach of [22] with the subset of APs that support the $k$th user. The signal received and processed at the CPU is expressed by

$$\tilde{x}_k = \left( \sum_{l=1}^L d_{kl} a_{kl}^* \mathbf{w}_{kl}^\mathsf{H} \mathbf{h}_{kl} \right) \sqrt{\eta_k} x_k + \tag{6}$$

$$\sum_{l=1}^L a_{kl}^* \left( \sum_{i=1, i \neq k}^K d_{kl} \mathbf{w}_{kl}^\mathsf{H} \mathbf{h}_{il} \sqrt{\eta_i} x_i \right) + \sum_{l=1}^L d_{kl} a_{kl}^* \mathbf{w}_{kl}^\mathsf{H} \mathbf{v}_l,$$

where $\mathbf{a}_k = [a_{k1} \dots a_{kL}]^\mathsf{T} \in \mathbb{C}^L$ is the complex LSFD coefficient for AP $l$ and user $k$ used to reduce the inter-user interference. The AP selection coefficient $d_{kl}$ reduces the load of the fronthaul link since only some APs take part in the signal detection. Under the scenario that the CPU does not have the CSI, we adapt the approach of [23], [19] to our system model which allows the computation of the achievable spectral efficiency using the *use-and-then-forget* bound

$$\text{SE}_k = (1 - \tau_p/\tau_c) \log_2(1 + \text{SINR}_k), \tag{7}$$

where the signal-to-interference-plus-noise ratio (SINR) of the $k$-th user is described by

$$\text{SINR}_k = \frac{\eta_k |\mathbf{a}_k^{\text{H}} \mathbf{g}_k|^2}{\mathbf{a}_k^{\text{H}} \left( \sum_{i=1}^{K} \eta_i \Upsilon_{ki}^{(1)} - \eta_k \mathbf{g}_k \mathbf{g}_k^{\text{H}} + \sigma_v^2 \Upsilon_k^{(2)} \right) \mathbf{a}_k} \quad (8)$$

where

$$\mathbf{g}_k = \left[ \mathbb{E} \left\{ d_{k1} \mathbf{w}_{k1}^{\text{H}} \mathbf{h}_{k1} \right\}, \ldots, \mathbb{E} \left\{ d_{kL} \mathbf{w}_{kL}^{\text{H}} \mathbf{h}_{kL} \right\} \right]^{\text{T}}, \quad (9)$$

$$\Upsilon_{ki}^{(1)} = \left[ \mathbb{E} \left\{ d_{kl} \mathbf{w}_{kl}^{\text{H}} \mathbf{h}_{il} d_{kj} \mathbf{h}_{ij}^{\text{H}} \mathbf{w}_{kj} \right\} : l, j = 1, \ldots, L \right], \quad (10)$$

$$\Upsilon_k^{(2)} = \text{diag} \left( \mathbb{E} \left\{ \| \mathbf{w}_{k1} d_{k1} \|^2 \right\}, \ldots, \mathbb{E} \left\{ \| \mathbf{w}_{kL} d_{kL} \|^2 \right\} \right). \quad (11)$$

and the expected values are with respect to the random parameters. For scalability, instead of taking into account all users in the network we consider in the optimal LSFD weight computation only the users that cause substantial interference to the $k$th user. Thus, as described in [14], [6], we have

$$\mathbf{a}_k = \left( \sum_{i \in \mathcal{B}_k} \eta_i \Upsilon_{ki}^{(1)} + \sigma_v^2 \Upsilon_k^{(2)} \right)^{-1} \mathbf{g}_k, \quad (12)$$

where $\mathcal{B}_k$ is the index set of users supported by the same APs as user $k$ (inclusive). Since the means of (10) and (11) are non-zero and deterministic they can be assumed known without loss of efficiency, enabling the computation of $\mathbf{a}_k$ and the decoding procedure in the CPU [5]. With the decentralized detection explained, we introduce in the next subsections the proposed LLR computation and AP selection procedures.

### A. Log-likelihood ratio computation

We propose the computation of local LLRs to perform AP selection. The main idea is to include each AP in an IDD scheme and use the reliability of the LLRs to improve the AP selection. An IDD scheme is used at the CPU to approach the performance of the maximum-likelihood (ML) detector [21], [24], [25]. In our IDD scheme, a soft MMSE detector processes extrinsic information provided by the channel decoder which incorporates soft information provided by the detector in (5). We adopted the soft MMSE detector and modified it to a distributed scheme [26] due to its suitability for processing LLRs in IDD schemes. Other detectors such as zero forcing and matched filter could be used at the expense of some performance degradation and various enhanced interference cancellation strategies [27], [28], [29], [30], [31], [32], [33], [34], [35], [36] could be incorporated to obtain performance benefits. The extrinsic information between the detector and decoder is exchanged in an iterative fashion until a maximum number of outer iterations is performed [31], [37]. In the proposed IDD scheme that is specifically designed for cell-free scenarios, we consider a low-density parity-check (LDPC) code due to its high efficiency and excellent performance and the LDPC decoder performs inner iterations. The extrinsic information produced by the linear MMSE detector is the difference of the soft-input and soft-output LLR values on the coded bits. In the $l$th AP, let $b_k(i)$ represent the $i$th bit of the modulated symbol $x_k$, transmitted by the $k$th user. Considering

$M_c$ as the modulation order ($i \in \{1, \ldots, M_c\}$), the extrinsic LLR value of the estimated bit $(b_k(i))$ is given by

$$\begin{aligned} \text{Lg}_{kl} \left( b_k(i) \right) &= \log \frac{P \left( b_k(i) = 1 | \hat{x}_{kl} \right)}{P \left( b_k(i) = 0 | \hat{x}_{kl} \right)} - \log \frac{P \left( b_k(i) = 1 \right)}{P \left( b_k(i) = 0 \right)} \\ &= \log \frac{\sum_{x_k \in \mathcal{A}_i^1} P \left( \hat{x}_{kl} | x_k \right) P \left( x_k \right)}{\sum_{x_k \in \mathcal{A}_i^0} P \left( \hat{x}_{kl} | x_k \right) P \left( x_k \right)} - \text{Lc}_{kl} \left( b_k(i) \right), \quad (13) \end{aligned}$$

where $\text{Lc}_{kl} \left( b_k(i) \right)$ is the extrinsic information of $b_k(i)$ computed by the LDPC decoder in the previous iteration between the detector and the LDPC decoder. $\mathcal{A}_i^1$ and $\mathcal{A}_i^0$ represent the set of $2^{M_c-1}$ hypotheses for which the $i$-th bit is 1 or 0, respectively. In the above computation of $\text{Lg}_{kl} \left( b_k(i) \right)$, the *a priori* probability $P(x_k)$ is given by

$$P(x_k) = \prod_{i=1}^{M_c} \left[ 1 + \exp \left( -\bar{b}_k(i) \, \text{Lc}_{kl} \left( b_k(i) \right) \right) \right]^{-1} \quad (14)$$

where $\bar{b}_k(i)$ denotes the state of the $i$th bit of the symbol of the $k$th user [21][1]. The likelihood function $P \left( \hat{x}_{kl} | x \right)$ is approximated by

$$P \left( \hat{x}_{kl} | x_k \right) \simeq \frac{1}{\pi \gamma_{kl}^2} \exp \left( -\frac{1}{\gamma_{kl}^2} | \hat{x}_{kl} - \alpha_{kl} x_k |^2 \right). \quad (15)$$

As expressed in (4), the output of the linear MMSE receive filter includes the desired symbol, residual interference and noise. As the inputs of the filter are corrupted versions of symbols that belong to a modulation scheme, the filter outputs are neither Gaussian nor i.i.d., which requires an approximation to compute the estimates $\alpha$ and $\gamma$ in (15). Therefore, we approximate $\hat{x}_{kl}$ by the output of an equivalent AWGN channel with $\hat{x}_{kl} = \alpha_{kl} x_k + z_k$, where $\alpha_{kl}$ is a scalar variable equivalent to the $k$-th UEs amplitude and $z_k \sim \mathcal{N}_c(0, \gamma_k^2)$. Since we have $\alpha_{kl} = \mathbb{E} \left[ x_k^* \hat{x}_{kl} \right]$ and $\gamma_{kl}^2 = \mathbb{E} \left[ | \hat{x}_{kl} - \alpha_{kl} x_l |^2 \right]$, the designer can obtain the estimates $\alpha_{kl}$ and $\gamma_{kl}$ via the corresponding sample averages over the frame transmission [27]. An alternative is to employ an approximation given by $\alpha_{kl} = \mathbf{w}_{kl}^{\text{H}} \hat{\mathbf{h}}_{kl}$, and $\gamma_{kl}^2 = \mathbb{E} \{ \hat{x}_{kl}^2 \} - \alpha_{kl}^2$ [38], [21]. Moreover, decoding performance benefits can be obtained by reweighting [39] and scheduling [40], [41], [42] techniques.

### B. AP selection procedure

In the cell-free scenario, the non-existence of cells and their boundaries leads to pilot contamination and the number of user connections of each AP is practically limited. Therefore, there is need for some coordination between APs and users for primary access. To begin the procedure of primary access, we determine the set of APs that could support each user. Then, we assume that each user will be served by at least a single AP with all its $N$ antenna elements as advocated in [5]. This means that $|\mathcal{D}_l| \leq \tau_p$ and $d_{kl} = 1$ if $k \in \mathcal{D}_l$ and $d_{kl} = 0$, otherwise. After the initial access that must enforce that every user can transmit its symbols to at least one AP, the APs build their lists of connections with users. Unlike the recent techniques that focus on using the LLSF [5], [19], [17] and LECG [12] criteria to carry out the initial access, in this work we use the LLRs. Differently from the recent study in [20] that employs LLRs, we do not consider massive machine-type communications, which requires activity detection, and we refine the distributed detection procedure and the cooperation between the APs and the CPU. The key idea is to compare the outputs of the IDD scheme of each supported UE in AP $l$ and only support the UEs with most reliable LLRs. After computing the local soft symbol estimates with (4), the IDD scheme starts and computes the LLRs until they reach the maximum number of outer iterations. With the LLRs, each AP computes the mean absolute value of the transmitted frame of each UE. Then, other means are computed, but in this case between each UE and its neighbours. The UEs that have their mean values of LLRs larger than those of the mean value of the group are assigned to the final list of that AP. Otherwise, it is not considered in the CPU detection step.

---

[1]Considering a QPSK modulation scheme, for the symbol $x_k = \sqrt{2}/2(1 - j)$, we have $b_k(1) = 1$, $b_k(2) = 0$, $\bar{b}_k(1) = +1$ and $\bar{b}_k(2) = -1$.

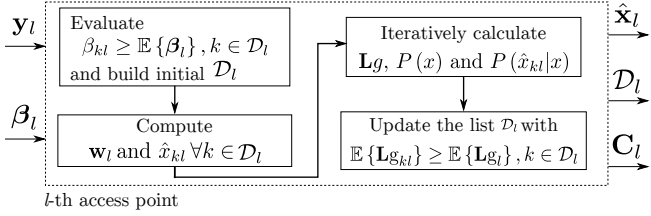

Fig. 1. Block diagram of the proposed LLR-M APs selection scheme.

The proposed AP selection algorithm, denoted as Log-Likelihood Mean (LLR-M), is illustrated in Fig. 1 and described in the following steps:

a) After the initial access and channel estimation, each AP uses the LLSF or LECG criterion to build its first version of $\mathcal{D}_l$. In the LLR-M algorithm, each AP includes the users index in $\mathcal{D}_l$ the LLSF coefficients larger than the mean of the coefficients of other users, i.e.,

$$\text{if } \beta_{kl} \geq \mathbb{E}\{\boldsymbol{\beta}_l\}, k \in \mathcal{D}_l \tag{16}$$

where $\boldsymbol{\beta}_l = [\beta_{1l}, \ldots, \beta_{jl}, \ldots, \beta_{Kl}], \forall j \neq k$;

b) Each user sends its frame to the APs that obtained a connection;

c) The $l$th AP receives the signals of all users that are included in $\mathcal{D}_l$ and carries out the symbol detection of each symbol interval, with (4) and (5);

d) The LLRs associated with each symbol of the transmit frame are obtained by the IDD scheme with (13), (14) and (15);

e) The $l$th AP makes a decision about whether it should detect the symbol of the $k$th user that will be processed at the CPU based on the reliability of the symbol. If the mean of the absolute value of the LLRs of the transmitted frame of the $k$th user is larger than the mean of the LLRs of the frames of all users in the set $\mathcal{D}_l$ then the user is included according to

$$\text{if } \mathbb{E}\{\mathbf{Lg}_{kl}\} \geq \mathbb{E}\{\mathbf{Lg}_l\}, k \in \mathcal{D}_l, \text{with} \tag{17}$$

$\mathbf{Lg}_l = [\mathbb{E}\{\mathbf{Lg}_{1l}\}, \cdots, \mathbb{E}\{\mathbf{Lg}_{jl}\}, \cdots, \mathbb{E}\{\mathbf{Lg}_{Kl}\}], j \neq k.$

f) Each set $\mathcal{D}_l$ is updated and the linear MMSE receive filters computed in (5) and the channel statistics are transmitted to the CPU via fronthaul links.

At the receiver, using the set $\mathcal{D}_l$ with the soft estimates of the symbols $\hat{\mathbf{x}}_l$ and the channel statistics, the CPU performs soft information processing and detection in (6) with the interference cancellation aided by the LSFD coefficient with (9)-(12). We remark that the extrinsic LLRs were adopted for AP selection because they provided the best performance among several approaches evaluated, which included the LLRs, LECG and LLSF. Note that the proposed LLR-M algorithm also uses the large-scale fading coefficients but in a different way to LLR-LLSF and LLR-LECG. LLR-M actually employs the mean of the LLRs and compares it with the mean of the large-scale coefficients of other users. The proposed LLR-M algorithm only conveys the channel parameters and local estimates of a reduced number of APs that have been selected. Assuming perfect synchronization of all users, the connection request of a new user results in extra waiting time for the new transmission round to be evaluated with all connected users to the $l$th AP. If the users in the list of active users still have symbol frames to transmit, then the new user will have its $\beta_{kl}$ and $\mathbb{E}\{\mathbf{Lg}_{kl}\}$ computed and compared to the current $\mathbb{E}\{\mathbf{Lg}_l\}$. On the other hand, if a user ends its transmission, it leaves the list and $\mathbb{E}\{\mathbf{Lg}_l\}$ is computed based on the users that are connected to AP $l$. After that, the IDD scheme carries out the iterative information processing using outer iterations between the detector and the decoder, which performs inner iterations.

## IV. COMPLEXITY AND SCALABILITY

In this section, we evaluate the time and space complexity, and scalability of the IDD and AP selection schemes. With the "Random" AP selection as a reference for worst performance and "All APs" as the benchmark for best performance where all users are supported by all APs, the LSF [5], [19], [17] and LECG [12] criteria were evaluated along with the proposed LLR-M and two LLR-based AP selection schemes initialized by the LLSF (LLR-LLSF) and LECG (LLR-LECG) criteria.

In terms of time complexity in floating point operations (FLOPS), we consider the cooperation level between APs and the CPU, the proposed IDD and AP schemes compute linear MMSE receive filters with a cost of $O(LN^3)$, whereas the centralized soft interference cancellation (IC) IDD scheme of [21] requires $O((LN)^3)$, the multi-branch decision feedback (MB-DF) scheme of [24] with $B$ branches needs $O(B(LN)^3)$ and the joint channel estimation and data detection (JED) scheme of [25] requires $O((2K)^3) + O(6K(LN)^2) + O(4KLNn) - O(2K) + 1$, where $n$ is the LLR block size. The LLR computation needs $2M_c$ to obtain (14), $2 \cdot 2^{M_c}$ to compute (13) and 4 FLOPS to evaluate (15). The computational cost of the LLR-based AP selection schemes is given by $(^K/_2)(|\mathcal{M}_k|^2 + |\mathcal{M}_k|) + (^1/_3)(|\mathcal{M}_k|^3 - |\mathcal{M}_k|) + |\mathcal{M}_k|^2 + 2(M_c + 2^{M_c}) + 4$, where $\mathcal{M}_k$ denotes the set of APs that serve the $k$th user.

In terms of space complexity in bytes, which is the amount of memory space required to solve an instance of the problem as a function of the input, the proposed IDD and AP selection schemes require $O(LN^2)$ per user for the covariance matrices, $O(N)$ for the received signal of each of the $L$ APs and $nM_c$ for the storage of blocks of $n$ LLRs, whereas the centralized IDD scheme of [21], the MB-DF [24] and the JED [25] detectors require $O((LN)^2)$ for the covariance matrices and a comparable space complexity to the proposed IDD scheme for the other tasks. Thus, the time and space complexities of the proposed IDD and AP selection schemes are lower than those in the literature because of their distributed nature and the fact that the covariance matrices stored have lower dimensions than those of the centralized approaches.

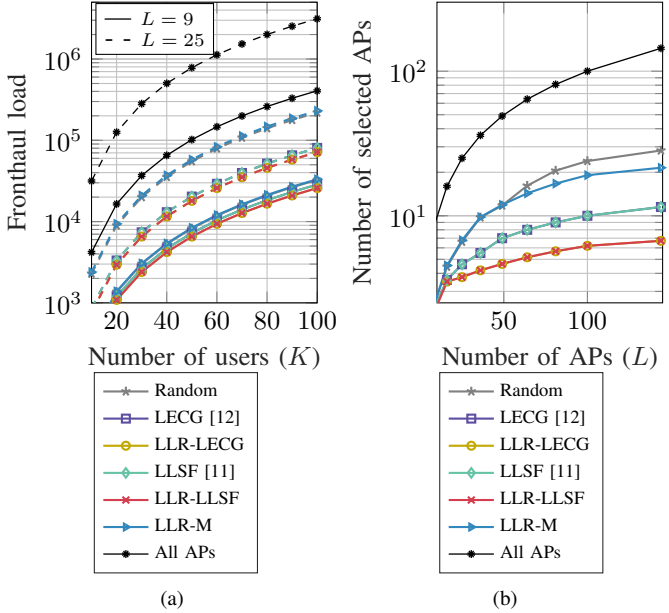

Fig. 2. (a) Fronthaul load in terms of complex scalars vs. different $K$ values for $N = 4$, $L = 9, 25$ and (b) Average number of selected APs per user vs. different network size.

The fronthaul load is expressed in terms of complex scalars and depends only on $K$ and $|\mathcal{M}_k|$, as expressed by [19], [6]

$$K|\mathcal{M}_k| + \left(|\mathcal{M}_k|^2 K^2 + K|\mathcal{M}_k|\right)/2. \quad (18)$$

Unlike previously reported AP selection schemes applied to cell-free systems, this work does not require matrix inversions, thus the time computational complexity and the fronthaul load of each AP selection scheme differs mainly due to the set $|\mathcal{M}_k|$ (which is directly related to $\mathcal{D}_l$). Fig. 2(a) shows the fronthaul load of different AP selection techniques with $L = 9$ and $L = 25$, which shows that despite the enlarged network, the required fronthaul signaling load remains approximately constant. We also show the number of selected APs versus the number of APs ($L$) in Fig. 2 (b), where $N = 4$ antennas per AP and $K = \lfloor L/4 \rfloor$ users. Fig. 2(b) shows that the LLR-based schemes initialized by LECG and LLSF metrics require less AP connections than the pure LECG and LLSF schemes, which is beneficial in terms of scalability. Furthermore, the proposed LLR-M technique requires higher fronthaul load than that of the approaches initialized by LLSF or LECG, but the gain in performance of LLR-M over the approaches initialized by LLSF or LECG is significant.

## V. NUMERICAL RESULTS

In this section we evaluate the proposed IDD and LLR-based AP selection schemes against existing solutions in terms of SE and BER obtained by numerical results. We also remark that the proposed IDD scheme is a modified and distributed version of the soft IC IDD scheme of [21] that has no AP selection. We investigate a scenario with $L$ APs equipped with uniform linear arrays with half-wavelength-spacing and $N$ antenna elements that support $K$ single-antenna users that are independently and uniformly distributed in an area of $1 \text{ Km}^2$. The difference in height between the APs and the users is set to 10 m, the frequency bandwidth centered over the carrier frequency of 2 GHz is chosen as 20 MHz, the power spectral density of the noise is 174 dBm/Hz and the noise figure is 5 dB. To compare the proposed techniques with the existing approaches, we use the spatial correlation matrix given by tables B.1.2.1-1 and B.1.2.1-4 in [43]: $\beta_{kl} = -30.5 - 36.7 \log_{10}\left(\delta_{kl}/1\,\text{m}\right) + S_{F_{kl}}$, where $\delta_{kl}$ is the distance between user $k$ and AP $l$ and $S_{F_{kl}} \sim \mathcal{N}_{\mathbb{C}}\left(0, 4^2\right)$ is the shadow fading and $\beta_{kl}$ is given in dB. Even though the propagation model in [43] was conceived for cellular systems, we modified it for the cell-free setting, as in [6], [15]. Furthermore, the spatial correlation is obtained using the Gaussian local scattering model with $15°$ angular standard deviation and the coherence frames have $\tau_c = 140$, where $\tau_u = 128$ is used for uplink transmissions. The proposed LLR-M technique for AP selection is based on the computation of LLRs, the channel coding scheme considered employ LDPC codes [44], [45] with rate $R_{\text{LDPC}} = 1/2$, and the modulation scheme is QPSK. In studies with IDD schemes and LDPC codes, half-rate LDPC codes are often adopted although other rates could be used as LDPC codes are quite flexible. For instance, a low-rate code such as $R_{\text{LDPC}} = 1/4$ would be able to lower even more the BER and enhance the interference mitigation capability although its efficiency would be lower. We have used 3 outer iterations between the detector and the decoder, and 20 inner iterations in the LDPC decoder. The average SNR of the $l$th AP is given by $\text{SNR}_l = \sum_{i=1}^{K} \beta_{il} \eta_i R_{\text{LDPC}}\left(1/\sigma_{v_{il}}^2\right)$. Since the proposed LLR-M technique for AP selection does not consider power control, we adopted a cost-effective technique for all approaches, namely, the fractional power control (FPC) [15], [19], in particular, the FPC of [15].

We assess the proposed and the existing IDD and AP selection schemes in the cell-free scenario of $L = 100$ APs, $K = 100$ users and $N = 4$ antennas, where all approaches perform the same soft iterative processing, detection and channel estimation procedures. For every independent trial of $10^4$ Monte Carlo runs, new user locations are generated. Fig. 3 (a) shows the cumulative distribution function (CDF) of the spectral efficiency per user, Fig. 3 (b) depicts the number of FLOPS versus $LN$. The CDF results indicate that the LLR-M AP selection approaches the performance of the system with all APs and outperforms the remaining AP selection techniques. In terms of complexity, the proposed distributed IDD scheme and AP selection is less complex than the soft IC [21], the MB-DF [24] and JED [25] schemes.

Fig. 4 shows the BER against the average signal-to-noise ratio (SNR). The results in Fig. 4 (a) show that the proposed IDD scheme with AP selection, the MB-DF [24] with 8 branches and the JED [25] IDD schemes have comparable performance when all APs are employed, whereas their performance gap increases slightly with the AP selection based on LLSF [11]. Nevertheless, the simplicity and lower cost of the proposed soft IC scheme motivated us to adapt it and incorporate an AP selection technique.

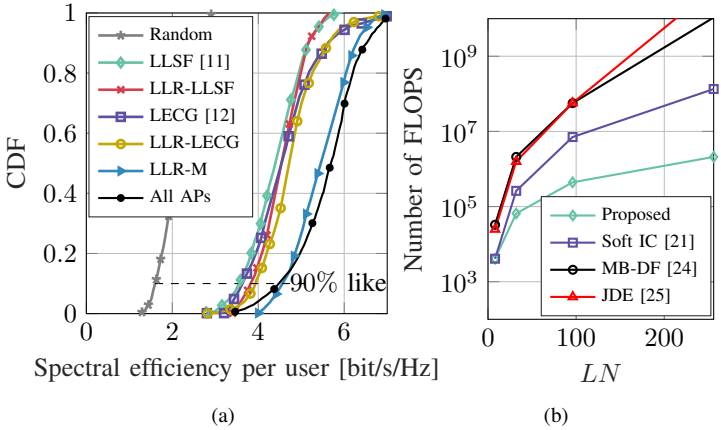

Fig. 3. (a) Spectral efficiency per user (bits/Hz) and (b) Complexity in FLOPs.

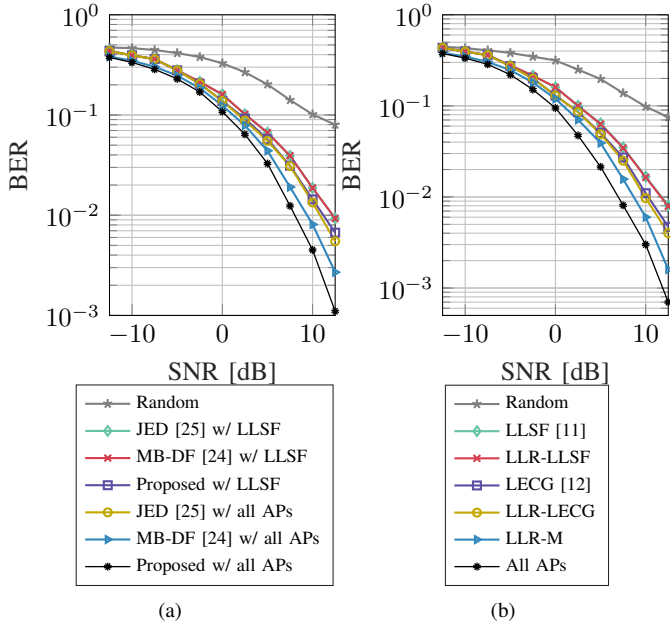

Fig. 4. (a) BER vs. SNR in dB for the different IDD schemes and (b) BER vs. SNR in dB for the different APs selection schemes.

The results in Fig. 4 (b) indicate that the LLR-based AP selection scheme initialized by the LLSF [11] and LECG [12] performs better than the standard LLSF [11] and LECG [12] approaches, but as shown in Fig. 2, requires almost the same fronthaul load and less connections to APs than the standard LLSF and LECG approaches. Conversely, the proposed LLR-M technique for AP selection outperforms the other approaches in terms of BER, as depicted in Fig. 4. In terms of average BER across all connected users, the results of the evaluated AP selection approaches indicate that the same performance hierarchy of Fig. 3 (a) is observed in Fig. 4.

## VI. CONCLUDING REMARKS

In this letter, we have proposed a distributed IDD scheme and an AP selection scheme for the uplink of a cell-free massive MIMO system. The proposed IDD scheme is highly effective for signal detection, resulting in excellent BER performance, whereas the proposed LLR-based AP selection tecnique considerably reduces the number of connected APs to the fronthaul links, reducing the fronthaul signaling load. By exploiting the reliability of the locally computed LLRs, the proposed IDD and AP selection schemes outperform existing techniques in terms of CDF and BER versus SNR results.